\documentclass[12pt]{article}
\usepackage{amssymb,amsmath}
\newcommand{\lla}[1]{\left\{#1\right\}}

\newcommand{\der}[2]{\frac{\partial #1}{\partial #2}}
\newcommand{\cit}[1]{(\ref{#1})}

\newcommand{\pare}[1]{\left(#1\right)}
\newcommand{\pb}[2]{\lla{{#1},{#2}}}
\newcommand{\db}[2]{\pb{#1}{#2}^*}
\newcommand{\spc}{\rule{0in}{5ex}}

\newcommand{\sub}[1]{_{\scriptstyle{#1}}}
\renewcommand{\sup}[1]{^{\scriptstyle{#1}}}

\newcommand{\func}[1]{\int d\sup{#1} x}

\begin{document}
\begin{center}
{\huge Equivalencia Can\'onica entre Teor\'{\i}as de 
esp\'{\i}n 1 Masivo}\\
\vskip 1cm
{\bf P\'{\i}o J. Arias${}^{a,b,1}$ y Jean C. P\'erez-Mosquera${}^{a,2}$}\\
${}^a${\it Grupo de Campos y Part\'{\i}culas, Dpto. de F\'{\i}sica, 
Facultad de Ciencias,\\
Universidad Central de Venezuela, AP47270, Caracas 1041-A, Venezuela}\\
\vskip 12pt
${}^b${\it Centro de Astrof\'{\i}sica Te\'orica, Facultad de Ciencias, 
Universidad de Los Andes, AP26, La Hechicerta, M\'erida 5101, Venezuela}\\
\vskip 12pt
${}^1${\tt parias@fisica.ciens.ucv.ve}, 
${}^2${\tt jcperez@fisica.ciens.ucv.ve}
\end{center}
\vskip 1cm
\begin{abstract}
Se considera el modelo de Cremmer-Scherck, generalizado, en contraposici\'on con el 
modelo de Proca en dimensiones mayores que 3+1. Se muestra que el modelo de Proca es
 una versi\'on  de la de Cremmer y Scherck con el calibre fijado, adem\'as se muestra 
la equivalencia can\'onica entre estos.
\end{abstract}

\begin{center}
{\bf{Abstract}}\\
\begin{narrower}
The model of Cremmer-Scherck and Proca are considered in dimensions greater than 3+1. It is 
obtained that the Proca model correspond to a gauged fixed version of the Cremmer-Scherck one, and we show their canonical equivalence.
\end{narrower}
\end{center}

\begin{section}{Introducci\'on}
En 3+1 dimensiones, la acci\'on del modelo de Proca es 
\begin{equation}
S=\int_Md\sup 4x\left[-\frac 14 F\sub{\mu\nu} F\sup{\mu\nu}-\frac{\mu\sup 2}2 A\sub\mu A\sup\mu\right],\label{eq:1}
\end{equation}
donde $F\sub{\mu\nu}=\partial\sub\mu A\sub\nu-\partial\sub\nu A\sub\mu$ es la intensidad de campo electromagn\'etico. 
Esta acci\'on no posee invariancias locales y describe part\'{\i}culas masivas de esp\'{\i}n 1.

Otra teor\'{\i}a vectorial que describe esp\'{\i}n $1$ masivo en $3+1$ es la de Cremmer y Scherk, cuyo funcional de acci\'on, invariante de calibre, es \cite{CS}
\begin{equation}\label{eq:2}
S\sub{CS}\sup 4=\int_M\!\!\! d\sup 4x\left[-\frac 14 F\sup{\mu\nu}F\sub{\mu\nu}-\frac 1{12\mu\sup 2}H\sup{\mu\nu\lambda}H\sub{\mu\nu\lambda}-\frac 14\epsilon\sup{\mu\nu\lambda\rho}B\sub{\mu\nu}F\sub{\lambda\rho}\right],
\end{equation}
donde $H\sub{\mu\nu\lambda}=\partial\sub\mu B\sub{\nu\lambda}+\partial\sub\lambda B\sub{\mu\nu}+\partial\sub\nu B\sub{\nu\lambda}$ es la intensidad del campo de Kalb-Ramond.

Se puede probar que ambas teor\'{\i}as son equivalentes en regiones del espacio-tiempo
 con topolog\'{\i}a trivial \cite{ABL,ARL}. En este trabajo veremos que si se 
generalizaran ambas teor\'{\i}as a $d+1$ dimensiones, usando el dual del potencial 
de Kalb-Ramond, la equivalencia persiste.
\end{section}
\begin{section}{Teor\'{\i}as Duales}
La acci\'on \eqref{eq:1}, puede ser escrita a primer orden si introducimos una 2-forma auxiliar $B=\frac 12B\sub{\mu\nu}dx\sup\mu dx\sup\nu$ \cite{FT}.
\begin{eqnarray}
S\sub P[A,B]&=&\int_M\left[\frac 14\epsilon\sup{\mu\nu\lambda\rho}F\sub{\mu\nu}B\sub{\lambda\rho}-\frac 14 B\sub{\mu\nu}B\sup{\mu\nu}-\frac{\mu\sup 2}2 A\sub\mu A\sup\mu \right],\nonumber\\
\spc
      &=&\frac 12\left(-(F,^\star\! B)+\frac 14 (B,B)+\mu\sup 2(A,A)\right)\label{eq:4},
\end{eqnarray}
donde
\begin{equation}
(\omega,\eta)\equiv\int_M\omega\wedge\sup\star\eta,
\end{equation}
y usamos la notacion de la referencia \cite{N} para el lenguaje de formas diferenciales.
Las ecuaciones de movimiento que se obtienen de $S\sub P$ son
\begin{eqnarray}
d\sup{\dagger}{}\sup\star B-\mu\sup 2A&=&0,\nonumber\\
\sup\star dA-B&=&0,\label{eq:6}
\end{eqnarray}
eliminando $B$ de \eqref{eq:6} tenemos
\begin{equation}
d\sup\dagger F-\mu\sup 2=0,
\end{equation}
las cuales son las ecuaciones de la teor\'{\i}a de Proca.

La acci\'on \eqref{eq:4} la generalizamos a $d+1$ dimensiones 
considerando a $B$ como una $d-1$ forma \cite{ARL} e introduciendo 
la 2-forma $T\equiv \quad \sup\star B$, en lugar de $B$, como variable de campo. 
De este modo obtenemos la acci\'on
\begin{equation}\label{eq:7}
S_P\sup\star[A,T]=-(F,T)-\frac 12(T,T)+\frac{\mu\sup 2}2(A,A),
\end{equation}
cuyas ecuaciones de movimiento  son
\begin{eqnarray}
-d\sup\dagger T+\mu\sup 2A&=&0,\nonumber\\
dA+T&=&0.\label{eq:8}
\end{eqnarray}
La teor\'{\i}a descrita por la acci\'on   \eqref{eq:7}, la llamaremos formulaci\'on dual de Proca.

Por otro lado, la acci\'on de Cremmer y Scherck
\begin{equation}\label{eq:9}
S_{CS}[A,B]=(F,\sup\star B)+\frac 1{2\mu\sup 2}(H,H)+\frac 12(F,F),
\end{equation}
puede ser generalizada a $d+1$ dimensiones de la misma manera que Proca \cite{ARL}. As\'{\i}, la formulaci\'on dual de la teor\'{\i}a de Cremmer y Scherck estar\'a descrita por la acci\'on
\begin{equation}\label{eq:10p}
S_{CS}\sup\star[A,T]=(F,T)-\frac 1{2\mu^2}(h,h)+\frac 12(F,F),
\end{equation}
donde $T=\sup\star B$ y $h=d\sup\dagger T=\partial\sup\nu T\sub{\mu\nu}dx\sup\mu$. Los t\'erminos 
cin\'eticos de Maxwell y Kalb-Ramond, estando solos, describen excitaciones sin masa. 
Ahora estando juntos con el t\'ermino de interacci\'on entre $A$ y $T$, que es de naturaleza 
topol\'ogica, describen una excitaci\'on masiva. Es por esto que nos referiremos al modelo de Cremmer y Scherck 
como Topol\'ogico Masivo (TM).

Las ecuaciones de movimiento asociadas a la acci\'on \eqref{eq:10p} son:
\begin{eqnarray}
d\sup\dagger(T+F)&=&0,\nonumber\\
d\pare{d\sup\dagger T-\mu\sup 2A}&=&0.\label{eq:10}
\end{eqnarray}

De las ecuaciones \eqref{eq:6}  y  \eqref{eq:10}  se observa que las soluciones de Proca son soluciones de la TM. Por otra parte, las ecuaciones de la TM son siempre satisfechas si $A$ es cerrada y $T$ es cocerrada. Sin embargo, en el caso de Proca, las ecuaciones \eqref{eq:6}  dicen que $T=0$ si $A$ es cerrada y $A=0$ si $T$ es cocerrada.

En consecuencia, existe un sector en el espacio de soluciones cl\'asicas libres de la TM que no est\'a presente en el de Proca, formado por todas las formas no nulas $A$ y $T$, cerradas (no exactas) y cocerradas (no coexactas) respectivamente. Este sector es de naturaleza topol\'ogica, puesto que depende de la estructura topol\'ogica de la variedad base. Este sector pertenece al espacio de soluciones cl\'asicas del modelo B.F., cuya acci\'on (dual) es,
\begin{equation}
S\sub{B.F.}=\frac 12 (F,T)=-\int d\sup dx\frac 12 T\sup{\mu\nu}F\sub{\mu\nu}.
\end{equation}

Las ecuaciones de la TM son localmente equivalentes a
\begin{eqnarray}
A-\frac 1{\mu\sup 2}d\sup\dag T&=&d\lambda\label{poinc1},\\
dA+ T&=&d\sup\dag\Lambda\label{poinc2}.
\end{eqnarray}
Esto tambi\'en ocurre en regiones donde el primero y el $(d-1)-$\'esimo grupo de 
cohomolog\'{\i}a son triviales. En virtud de la invariancia de calibre de la TM, 
podemos escoger un calibre que absorba a $\lambda$ y $\Lambda$, de modo que 
$\lambda=0$ y $\Lambda=0$, obteni\'endose as\'{\i} las ecuaciones de Proca, 
luego de esta fijaci\'on. Vemos, entonces que el modelo de Proca surge luego de 
una fijaci\'on de calibre en la TM.


\end{section}
\begin{section}{Equivalencia Can\'onica}
Como se vi\'o en la secci\'on anterior, la acci\'on de Proca se puede escribir en 
$d+1$ dimensiones en t\'erminos de la 2-forma $T=^\star\!\! B$ de la siguiente manera
\begin{eqnarray}
S\sub P\sup\star=&&-(F,T)-\frac 12(T,T)+\frac{\mu\sup 2}2(A,A),  \label{pdenformas}\\
S\sub P\sup\star=&&\frac 12\func{D}\left[T\sup{\mu\nu}F\sub{\mu\nu}+\frac 12T\sup{\mu\nu}T\sub{\mu\nu}-\mu\sup 2 A\sub\mu A\sup\mu\right]\label{pdencomponentes}.
\end{eqnarray}
Para construir el hamiltoniano hacemos la descomposici\'on $d+1$ de la acci\'on \cit{pdencomponentes}, i.e, 
\begin{eqnarray}
S\sub P\sup\star=\func{D}\left[\dot T\sub{0i}A\sub i+
T\sub{0i}\partial\sub i A\sub 0+\frac 12 T\sub{ij}F\sub{ij}
-\frac 12 T\sub{0i}T\sub{0i} +\frac 14 T\sub{ij}T\sub{ij}\right.\nonumber\\
\left.+\frac{\mu\sup 2}2 A\sub 0 A\sub 0
-\frac{\mu\sup 2}2 A\sub i A\sub i\right],\label{accionend+1}
\end{eqnarray}
donde hemos integrado por partes en el tiempo\footnote{Supondremos en todo el trabajo que los campos se anulan en la frontera de la regi\'on de integraci\'on, por lo tanto, los t\'erminos de borde no contribuyen.} para tomar como termino cin\'etico a $\dot T\sub{0i}A\sub i$ en lugar de $T\sub{0i}\dot A\sub i$. De la ecuaci\'on \eqref{accionend+1} vemos  que las variables $A\sub 0$ y $T\sub{ij}$ pueden ser consideradas variables no din\'amicas, puesto sus velocidades no aparecen en la acci\'on.

Definimos los momentos can\'onicos conjugados a los campos
\begin{eqnarray}
\Pi\sub i&=&\der{\mathcal L}{\dot A\sub i}=0,\\
\spc
p\sub i&=&\der{\mathcal L}{\dot T\sub{0i}}=A\sub i,
\end{eqnarray}
de donde se observa claremente que no se pueden despejar las velocidades, y en 
consecuencia tenemos el conjunto $\Psi\sub a=(\psi\sub i,\varphi\sub j)$ de $2d$ v\'{\i}nculos primarios
\begin{eqnarray}
\psi\sub i&=&\Pi\sub i\label{v1},\\
\varphi\sub i&=&p\sub i-A\sub i\label{v2}.
\end{eqnarray}
Estos v\'{\i}nculos definen la superficie $\Gamma_1$ de los v\'{\i}nculos primarios. El hamiltoniano sobre $\Gamma_1$ es
\begin{equation}
       H\sub P=\int d\sup d\vec x\left[\frac{\mu\sup 2}2 A\sub i A\sub i+\frac 12 T\sub{0i}T\sub{0i}+A\sub 0\left(\partial\sub i T\sub{0i}-\frac{\mu\sup 2}2 A\sub 0\right)-\frac 12 T\sub{ij}\left(F\sub{ij}+\frac 12 T\sub{ij}\right)\right].
\end{equation}

Pidiendo ahora que las variaciones de $H\sub P$ respecto de las variables no din\'amicas se anulen 
\begin{eqnarray}
\frac{\delta H\sub P}{\delta T\sub{ij}}&=&-\frac 12\left(F\sub{ij}+T\sub{ij}\right) =0, \\
\frac{\delta H\sub P}{\delta A\sub 0}&=&\partial\sub i T\sub{0i}-\mu\sup 2 A\sub 0=0, 
\end{eqnarray}
se obtienen expresiones de las cuales se pueden despejar $A\sub 0$ y $T\sub{ij}$ en funci\'on de las variables can\'onicas
\begin{eqnarray}
T\sub{ij}&=&-F\sub{ij},\label{nocan1}\\
A\sub 0&=&\frac 1{\mu\sup 2}\partial\sub i T\sub{0i}.\label{nocan2}
\end{eqnarray}
Sutituyendo \eqref{nocan1} y \eqref{nocan2} en $H\sub P$, obtenemos 
\begin{equation}
H\sub P=\int d\sup d\vec x\left[\frac{\mu\sup 2}2 A\sub i A\sub i+\frac 14 F\sub{ij} F\sub{ij}+\frac 1{2\mu\sup 2}{\left(\partial\sub i T\sub{0i}\right)}\sup 2+\frac 12 T\sub{0i} T\sub{0i}\right]\label{hproca}.
\end{equation}

La preservaci\'on en el tiempo de los v\'{\i}nculos nos conduce a ecuaciones que 
permiten despejar completamente los multiplicadores, de modo que el conjunto de 
v\'{\i}nculos $\psi_a$ es de segunda clase. Los corchetes de Dirac entre los campos 
$A\sub i$ y $T\sub{0i}$ resultan ser
\begin{equation}
\db{A\sub i(\vec x)}{T\sub{0j}(\vec y)}=-\delta\sub{ij}\delta\sup d(\vec x-\vec y),
\end{equation}
 
La acci\'on de Cremmer y Scherk en su formulaci\'on dual, se escribe en el lenguaje de formas de la siguiente manera
\begin{eqnarray}
S\sub{CS}\sup\star&=&(F,T)+\frac 12(F,F)-\frac 1{2\mu\sup 2}(h,h)\label{atmenformas}\\
&=&\int d\sup Dx\left[-\frac 14 F\sub{\mu\nu}F\sup{\mu\nu}-\frac 12 T\sub{\mu\nu}F\sup{\mu\nu}+\frac 1{2\mu\sup 2}h\sub\mu h\sup\mu\right]\label{atmencomponentes}. 
\end{eqnarray}

Para hacer el an\'alisis can\'onico procedemos a realizar la descomposici\'on $d+1$ de la acci\'on \eqref{atmencomponentes}, i.e.,
\begin{eqnarray}
S\sub{CS}=\int d\sup Dx\left[\frac 12 F\sub{0i}F\sub{0i}
-\frac 14 F\sub{ij}F\sub{ij}-\frac 1{2\mu\sup 2}h\sub 0 h\sub 0
+\frac 1{2\mu\sup 2}h\sub i h\sub i\right. \nonumber\\
\left.+T\sub{0i}F\sub{0i}-\frac 12 T\sub{ij}F\sub{ij}\right],
\end{eqnarray}
de aqu\'{\i} se observa que $A\sub 0$ y $T\sub{ij}$ se pueden considerar no din\'amicas, por la ausencia de sus velocidades en la acci\'on. Partimos de la definici\'on de los momentos can\'onicos conjugados a los campos
\begin{eqnarray}
\Pi\sub i=\der{\mathcal L}{\dot A\sub i}&=&F\sub{0i}+T\sub{0i}=\dot A\sub i-\partial\sub i A\sub 0+T\sub{0i},\\
\spc
p\sub i=\der{\mathcal L}{\dot T\sub{0i}}&=&\frac 1{\mu\sup 2}H\sub i=\frac 1{\mu\sup 2}\left(\dot T\sub{0i}+\partial\sub j T\sub{ij}\right).
\end{eqnarray}
de donde  se pueden despejar completamente las velocidades
\begin{eqnarray}
\dot A\sub i&=&\Pi\sub i+\partial\sub i A\sub 0-T\sub{0i},\\
\spc
\dot T\sub{0i}&=&\mu\sup 2 p\sub i-\partial\sub j T\sub{ij}.
\end{eqnarray}

Ahora procedemos a construir el Hamiltoniano,
\begin{eqnarray}
H&=&\func{d}\left[\frac 12\Pi\sub i\Pi\sub i+\frac {\mu\sup 2}2 p\sub i p\sub i-\Pi\sub{i}T\sub{0i}+\frac 12 T\sub{0i}T\sub{0i}+\frac 14 F\sub{ij}F\sub{ij}\right.\nonumber\\
\spc
 & &\qquad\qquad\qquad\quad\left.+\frac 1{2\mu\sup 2}H\sub 0 H\sub 0+\frac 12 T\sub{ij}\left(F\sub{ij}-\tilde F\sub{ij}\right)+A\sub 0\left(-\partial\sub i\Pi\sub i\right)\right],
\end{eqnarray}
donde $\tilde F\sub{ij}=\partial\sub i p\sub j-\partial\sub j p\sub i$.

Exigiendo que las variaciones de $H$ respecto de los campos no din\'amicos se anulen obtenemos el conjunto de v\'{\i}nculos $\Theta\sub a=(\theta,\theta\sub{ij})$, donde
\begin{eqnarray}
\theta=&&-\partial\sub i\Pi\sub i=0\label{vtm1}\\
\spc
\theta\sub{ij}=&&\frac 12\left(F\sub{ij}-\tilde F\sub{ij}\label{vtm2}\right).
\end{eqnarray}

Se puede ver que el conjunto $\Theta_a$ es de primera clase y las transformaciones de calibre generadas por estos son
\begin{eqnarray}
\delta A\sub i(x)=&&\partial\sub i\xi\\
\delta T\sub{0i}(x)=&&\partial\sub j\xi\sub{ji}
\end{eqnarray}

El Hamiltoniano que genera la din\'amica ser\'a
\begin{equation}
\tilde H\sub{TM}=\func{d}\left[\frac 12\Pi\sub i\Pi\sub i+\frac {\mu\sup 2}2 
p\sub i p\sub i-\Pi\sub iT\sub{0i}+\frac 12 T\sub{0i}T\sub{0i}+
\frac 14 F\sub{ij}F\sub{ij}+\frac 1{2\mu\sup 2}h\sub 0 h\sub 0+\lambda\sup a\Theta\sub a\right],
\end{equation}
donde $\lambda\sup a=(A\sub 0,T\sub{ij})$ son los multiplicadores indeterminados  de Lagrange asociados a los v\'{\i}nculos $\theta$ y $\theta\sub{ij}$.


Vamos ahora a probar la equivalencia antes mencionada desde el punto de vista hamiltoniano. 

El hamiltoniano de la teor\'{\i}a de Proca no invariante de calibre en $d+1$ dimensiones, una vez eliminadas las variables no din\'amicas, obtenido anteriormente, es
\begin{equation}
H\sub P=\int d\sup 3\vec x \left[\frac 14 F\sub{ij}F\sub{ij}+\frac 1{2\mu\sup 2}\left(\partial\sub i T\sub{0i}\right)\sup 2+\frac 12 T\sub{0i}T\sub{0i}+\frac{\mu\sup 2}2 A\sub i A\sub i\right],
\end{equation}
sujeto a los v\'{\i}nculos de segunda clase $\Psi\sub a=(\psi\sub i,\varphi\sub j)$
\begin{eqnarray}
\psi\sub i=\Pi\sub i\label{vinculos1},\\
\spc
\varphi\sub i=p\sub i-A\sub i\label{vinculos2}.
\end{eqnarray} 
Los corchetes de Poisson no nulos entre los v\'{\i}nculos son
\begin{equation}
\pb{\psi\sub i(x)}{\varphi\sub j(y)}=\delta_{ij}\delta\sup d(x-y).
\end{equation}
Podemos interpretar a la mitad de estos v\'{\i}nculos como de primera clase y 
la otra mitad como fijaciones de calibre. En lugar del conjunto $\Psi_a$ tomaremos 
el conjunto $\gamma_a=(\Theta_a,\chi_a)$, siendo 
$\chi_a=({\psi^{T}}_i,-\partial_i\varphi\sub i)$, respectivamente, la parte 
transversa de $\psi\sub i$ y la parte longitudinal de $\varphi\sub i$. Este nuevo 
conjunto de v\'{\i}nculos es completamente equivalente al primero en regiones del 
espacio con topolog\'{\i}a trivial. Tenemos entonces que el conjunto de  v\'{\i}nculos 
de la teor\'{\i}a de Proca se puede separar en los de la de Cremmer y Scherck m\'as 
otra parte que se pueden interpretar como fijaciones. Entonces buscamos un 
Hamiltoniano invariante de calibre de la forma \cite{ARL,GRS}
\begin{equation}
\tilde H\sub P=H\sub P+\int d\sup 3\vec x\left[\alpha\sub a(\vec x)\Theta\sub a(\vec x)+\beta\sub a(\vec x)\Phi\sub a(\vec x)+\int d\sup 3\vec y\beta\sub{ab}(\vec x,\vec y)\Phi\sub a(\vec x)\Phi\sub b(\vec y)\right]\label{hinvdecalibre}
\end{equation}
que difiere del de Proca por combinaciones de los v\'{\i}nculos.
Para hallar los coeficientes $\alpha 's$ y $\beta 's$ exijimos que los corchetes de 
este hamiltoniano con los v\'{\i}nculos de primera clase sean debilmente cero. 
Esto nos va a conducir a una familia de Hamiltonianos invariantes de calibre, 
puesto que los $\alpha$'s no quedar\'an determinados por la condici\'on antes 
mencionada. Sim embargo, podemos probar que el Hamiltoniano de $S\sub{CS}$ es uno 
de esta familia, simplemente observando que
\begin{eqnarray}
H\sub{TM}\sup{d+1}-H\sub P\sup{d+1}&=&\frac 12\Pi\sub i\Pi\sub i+\frac{\mu\sup 2}2p\sub i p\sub i+T\sub{0i}\Pi\sub i+\frac{\mu\sup 2}2A\sub i A\sub i,\\
                                   &=&\frac 12\psi\sub i\psi\sub i+T\sub{0i}\psi\sub i+\frac{\mu\sup 2}2\varphi\sub i\varphi\sub i+\mu\sup 2 A\sub i\varphi\sub i.
\end{eqnarray}
Con esto, queda probada la equivalencia can\'onica, que como dijimos est\'a sujeto 
a la cohomolog\'{\i}a de la variedad base. En otro direcci\'on puede verse que la funci\'on 
de partci\'on de ambos modelos difieren en un factor que correspon a la finci\'on de partici\'on de la acci\'on $S\sub {BF}$. 
La presencia de este factor se har\'a sentir en el momento de calcular valores esperados de objetos sensibles a la 
topolog\'{\i}a del espacio tiempo. En este sentido la descripci\'on a trav\'es de $S\sub P$ o $S\sub {CS}$ podr\'an ser distintas.
\end{section}

\end{document}